\documentclass[12pt,a4paper]{article}
\usepackage{epsfig}
\begin{document}
\mbox{\ }
\vskip -1.5cm
\centerline{\Large EUROPEAN ORGANIZATION FOR NUCLEAR RESEARCH}
\vskip 1cm
\begin{flushright}
CERN--EP/2000--119\\
8 September 2000
\end{flushright} 
\vskip 3.5cm
\Large
\centerline{\bf Study of the CP asymmetry of} 
\vskip 2mm
\centerline{\bf
  {\boldmath $\rm B^0 \rightarrow J/\psi\,K^0_S$} decays in ALEPH}
\normalsize
\vskip 2cm
\centerline{The ALEPH Collaboration\footnote{See the following pages for
    the list of authors.}}
\vskip 3cm

\begin{abstract}
The decay $\rm B^0 \rightarrow J/\psi\,K^0_S$ is reconstructed with $\rm
J/\psi \rightarrow e^+e^-$ or $\mu^+\mu^-$ and $\rm K^0_S \rightarrow
\pi^+\pi^-$. From the full ALEPH dataset at LEP1 of about 4 million
hadronic Z decays, 23~candidates are selected with an estimated purity of
71\%.  They are used to measure the CP asymmetry of this
decay, given by $\sin 2\beta$ in the Standard Model, with the result $\sin
2\beta = 0.84\ ^{+0.82}_{-1.04}\pm 0.16$.  This is combined with existing
measurements from other experiments, and increases the confidence level
that CP violation has been observed in this channel to 98\%.
\end{abstract}

\vfill
\centerline{\it Submitted to Physics Letters}
\vskip 1cm
\newpage
\pagestyle{empty}
\newpage
\small
%
\newlength{\saveparskip}
\newlength{\savetextheight}
\newlength{\savetopmargin}
\newlength{\savetextwidth}
\newlength{\saveoddsidemargin}
\newlength{\savetopsep}
\setlength{\saveparskip}{\parskip}
\setlength{\savetextheight}{\textheight}
\setlength{\savetopmargin}{\topmargin}
\setlength{\savetextwidth}{\textwidth}
\setlength{\saveoddsidemargin}{\oddsidemargin}
\setlength{\savetopsep}{\topsep}
%
%
\setlength{\parskip}{0.0cm}
\setlength{\textheight}{25.0cm}
\setlength{\topmargin}{-1.5cm}
\setlength{\textwidth}{16 cm}
\setlength{\oddsidemargin}{-0.0cm}
\setlength{\topsep}{1mm}
\pretolerance=10000
\centerline{\large\bf The ALEPH Collaboration}
\footnotesize
\vspace{0.5cm}
{\raggedbottom
\begin{sloppypar}
\samepage\noindent
R.~Barate,
D.~Decamp,
P.~Ghez,
C.~Goy,
\mbox{J.-P.~Lees},
E.~Merle,
\mbox{M.-N.~Minard},
B.~Pietrzyk
\nopagebreak
\begin{center}
\parbox{15.5cm}{\sl\samepage
Laboratoire de Physique des Particules (LAPP), IN$^{2}$P$^{3}$-CNRS,
F-74019 Annecy-le-Vieux Cedex, France}
\end{center}\end{sloppypar}
\vspace{2mm}
\begin{sloppypar}
\noindent
S.~Bravo,
M.P.~Casado,
M.~Chmeissani,
J.M.~Crespo,
E.~Fernandez,
\mbox{M.~Fernandez-Bosman},
Ll.~Garrido,$^{15}$
E.~Graug\'{e}s,
M.~Martinez,
G.~Merino,
R.~Miquel,
Ll.M.~Mir,
A.~Pacheco,
H.~Ruiz
\nopagebreak
\begin{center}
\parbox{15.5cm}{\sl\samepage
Institut de F\'{i}sica d'Altes Energies, Universitat Aut\`{o}noma
de Barcelona, E-08193 Bellaterra (Barcelona), Spain$^{7}$}
\end{center}\end{sloppypar}
\vspace{2mm}
\begin{sloppypar}
\noindent
A.~Colaleo,
D.~Creanza,
M.~de~Palma,
G.~Iaselli,
G.~Maggi,
M.~Maggi,$^{1}$
S.~Nuzzo,
A.~Ranieri,
G.~Raso,$^{23}$
F.~Ruggieri,
G.~Selvaggi,
L.~Silvestris,
P.~Tempesta,
A.~Tricomi,$^{3}$
G.~Zito
\nopagebreak
\begin{center}
\parbox{15.5cm}{\sl\samepage
Dipartimento di Fisica, INFN Sezione di Bari, I-70126
Bari, Italy}
\end{center}\end{sloppypar}
\vspace{2mm}
\begin{sloppypar}
\noindent
X.~Huang,
J.~Lin,
Q. Ouyang,
T.~Wang,
Y.~Xie,
R.~Xu,
S.~Xue,
J.~Zhang,
L.~Zhang,
W.~Zhao
\nopagebreak
\begin{center}
\parbox{15.5cm}{\sl\samepage
Institute of High Energy Physics, Academia Sinica, Beijing, The People's
Republic of China$^{8}$}
\end{center}\end{sloppypar}
\vspace{2mm}
\begin{sloppypar}
\noindent
D.~Abbaneo,
G.~Boix,$^{6}$
O.~Buchm\"uller,
M.~Cattaneo,
F.~Cerutti,
G.~Dissertori,
H.~Drevermann,
R.W.~Forty,
M.~Frank,
T.C.~Greening,
J.B.~Hansen,
J.~Harvey,
P.~Janot,
B.~Jost,
I.~Lehraus,
P.~Mato,
A.~Minten,
A.~Moutoussi,
F.~Ranjard,
L.~Rolandi,
D.~Schlatter,
M.~Schmitt,$^{20}$
O.~Schneider,$^{2}$
P.~Spagnolo,
W.~Tejessy,
F.~Teubert,
E.~Tournefier,
A.E.~Wright
\nopagebreak
\begin{center}
\parbox{15.5cm}{\sl\samepage
European Laboratory for Particle Physics (CERN), CH-1211 Geneva 23,
Switzerland}
\end{center}\end{sloppypar}
\vspace{2mm}
\begin{sloppypar}
\noindent
Z.~Ajaltouni,
F.~Badaud,
G.~Chazelle,
O.~Deschamps,
A.~Falvard,
P.~Gay,
C.~Guicheney,
P.~Henrard,
J.~Jousset,
B.~Michel,
S.~Monteil,
\mbox{J-C.~Montret},
D.~Pallin,
P.~Perret,
F.~Podlyski
\nopagebreak
\begin{center}
\parbox{15.5cm}{\sl\samepage
Laboratoire de Physique Corpusculaire, Universit\'e Blaise Pascal,
IN$^{2}$P$^{3}$-CNRS, Clermont-Ferrand, F-63177 Aubi\`{e}re, France}
\end{center}\end{sloppypar}
\vspace{2mm}
\begin{sloppypar}
\noindent
J.D.~Hansen,
J.R.~Hansen,
P.H.~Hansen,
B.S.~Nilsson,
B.A.~Petersen,
A.~W\"a\"an\"anen
\begin{center}
\parbox{15.5cm}{\sl\samepage
Niels Bohr Institute, DK-2100 Copenhagen, Denmark$^{9}$}
\end{center}\end{sloppypar}
\vspace{2mm}
\begin{sloppypar}
\noindent
G.~Daskalakis,
A.~Kyriakis,
C.~Markou,
E.~Simopoulou,
A.~Vayaki
\nopagebreak
\begin{center}
\parbox{15.5cm}{\sl\samepage
Nuclear Research Center Demokritos (NRCD), GR-15310 Attiki, Greece}
\end{center}\end{sloppypar}
\vspace{2mm}
\begin{sloppypar}
\noindent
A.~Blondel,$^{12}$
G.~Bonneaud,
\mbox{J.-C.~Brient},
A.~Roug\'{e},
M.~Rumpf,
M.~Swynghedauw,
M.~Verderi,
\linebreak
H.~Videau
\nopagebreak
\begin{center}
\parbox{15.5cm}{\sl\samepage
Laboratoire de Physique Nucl\'eaire et des Hautes Energies, Ecole
Polytechnique, IN$^{2}$P$^{3}$-CNRS, \mbox{F-91128} Palaiseau Cedex, France}
\end{center}\end{sloppypar}
\vspace{2mm}
\begin{sloppypar}
\noindent
E.~Focardi,
G.~Parrini,
K.~Zachariadou
\nopagebreak
\begin{center}
\parbox{15.5cm}{\sl\samepage
Dipartimento di Fisica, Universit\`a di Firenze, INFN Sezione di Firenze,
I-50125 Firenze, Italy}
\end{center}\end{sloppypar}
\vspace{2mm}
\begin{sloppypar}
\noindent
A.~Antonelli,
M.~Antonelli,
G.~Bencivenni,
G.~Bologna,$^{4}$
F.~Bossi,
P.~Campana,
G.~Capon,
V.~Chiarella,
P.~Laurelli,
G.~Mannocchi,$^{5}$
F.~Murtas,
G.P.~Murtas,
L.~Passalacqua,
\mbox{M.~Pepe-Altarelli}$^{24}$
\nopagebreak
\begin{center}
\parbox{15.5cm}{\sl\samepage
Laboratori Nazionali dell'INFN (LNF-INFN), I-00044 Frascati, Italy}
\end{center}\end{sloppypar}
\vspace{2mm}
\begin{sloppypar}
\noindent
A.W. Halley,
J.G.~Lynch,
P.~Negus,
V.~O'Shea,
C.~Raine,
\mbox{P.~Teixeira-Dias},
A.S.~Thompson
\nopagebreak
\begin{center}
\parbox{15.5cm}{\sl\samepage
Department of Physics and Astronomy, University of Glasgow, Glasgow G12
8QQ,United Kingdom$^{10}$}
\end{center}\end{sloppypar}
\vspace{2mm}
\begin{sloppypar}
\noindent
R.~Cavanaugh,
S.~Dhamotharan,
C.~Geweniger,$^{1}$
P.~Hanke,
G.~Hansper,
V.~Hepp,
E.E.~Kluge,
A.~Putzer,
J.~Sommer,
K.~Tittel,
S.~Werner,$^{19}$
M.~Wunsch$^{19}$
\nopagebreak
\begin{center}
\parbox{15.5cm}{\sl\samepage
Kirchhoff-Institut f\"r Physik, Universit\"at Heidelberg, D-69120
Heidelberg, Germany$^{16}$}
\end{center}\end{sloppypar}
\vspace{2mm}
\begin{sloppypar}
\noindent
R.~Beuselinck,
D.M.~Binnie,
W.~Cameron,
P.J.~Dornan,
M.~Girone,
N.~Marinelli,
J.K.~Sedgbeer,
J.C.~Thompson,$^{14}$
E.~Thomson$^{22}$
\nopagebreak
\begin{center}
\parbox{15.5cm}{\sl\samepage
Department of Physics, Imperial College, London SW7 2BZ,
United Kingdom$^{10}$}
\end{center}\end{sloppypar}
\vspace{2mm}
\begin{sloppypar}
\noindent
V.M.~Ghete,
P.~Girtler,
E.~Kneringer,
D.~Kuhn,
G.~Rudolph
\nopagebreak
\begin{center}
\parbox{15.5cm}{\sl\samepage
Institut f\"ur Experimentalphysik, Universit\"at Innsbruck, A-6020
Innsbruck, Austria$^{18}$}
\end{center}\end{sloppypar}
\vspace{2mm}
\begin{sloppypar}
\noindent
C.K.~Bowdery,
P.G.~Buck,
A.J.~Finch,
F.~Foster,
G.~Hughes,
R.W.L.~Jones,
N.A.~Robertson
\nopagebreak
\begin{center}
\parbox{15.5cm}{\sl\samepage
Department of Physics, University of Lancaster, Lancaster LA1 4YB,
United Kingdom$^{10}$}
\end{center}\end{sloppypar}
\vspace{2mm}
\begin{sloppypar}
\noindent
I.~Giehl,
K.~Jakobs,
K.~Kleinknecht,
G.~Quast,$^{1}$
B.~Renk,
E.~Rohne,
\mbox{H.-G.~Sander},
H.~Wachsmuth,
C.~Zeitnitz
\nopagebreak
\begin{center}
\parbox{15.5cm}{\sl\samepage
Institut f\"ur Physik, Universit\"at Mainz, D-55099 Mainz, Germany$^{16}$}
\end{center}\end{sloppypar}
\vspace{2mm}
\begin{sloppypar}
\noindent
A.~Bonissent,
J.~Carr,
P.~Coyle,
O.~Leroy,
P.~Payre,
D.~Rousseau,
M.~Talby
\nopagebreak
\begin{center}
\parbox{15.5cm}{\sl\samepage
Centre de Physique des Particules, Universit\'e de la M\'editerran\'ee,
IN$^{2}$P$^{3}$-CNRS, F-13288 Marseille, France}
\end{center}\end{sloppypar}
\vspace{2mm}
\begin{sloppypar}
\noindent
M.~Aleppo,
F.~Ragusa
\nopagebreak
\begin{center}
\parbox{15.5cm}{\sl\samepage
Dipartimento di Fisica, Universit\`a di Milano e INFN Sezione di Milano,
I-20133 Milano, Italy}
\end{center}\end{sloppypar}
\vspace{2mm}
\begin{sloppypar}
\noindent
H.~Dietl,
G.~Ganis,
A.~Heister,
K.~H\"uttmann,
G.~L\"utjens,
C.~Mannert,
W.~M\"anner,
\mbox{H.-G.~Moser},
S.~Schael,
R.~Settles,$^{1}$
H.~Stenzel,
W.~Wiedenmann,
G.~Wolf
\nopagebreak
\begin{center}
\parbox{15.5cm}{\sl\samepage
Max-Planck-Institut f\"ur Physik, Werner-Heisenberg-Institut,
D-80805 M\"unchen, Germany\footnotemark[16]}
\end{center}\end{sloppypar}
\vspace{2mm}
\begin{sloppypar}
\noindent
P.~Azzurri,
J.~Boucrot,$^{1}$
O.~Callot,
S.~Chen,
A.~Cordier,
M.~Davier,
L.~Duflot,
\mbox{J.-F.~Grivaz},
Ph.~Heusse,
A.~Jacholkowska,$^{1}$
F.~Le~Diberder,
J.~Lefran\c{c}ois,
\mbox{A.-M.~Lutz},
\mbox{M.-H.~Schune},
\mbox{J.-J.~Veillet},
I.~Videau,
C.~Yuan,
D.~Zerwas
\nopagebreak
\begin{center}
\parbox{15.5cm}{\sl\samepage
Laboratoire de l'Acc\'el\'erateur Lin\'eaire, Universit\'e de Paris-Sud,
IN$^{2}$P$^{3}$-CNRS, F-91898 Orsay Cedex, France}
\end{center}\end{sloppypar}
\vspace{2mm}
\begin{sloppypar}
\noindent
G.~Bagliesi,
T.~Boccali,
G.~Calderini,
V.~Ciulli,
L.~Fo\`{a},
A.~Giassi,
F.~Ligabue,
A.~Messineo,
F.~Palla,$^{1}$
G.~Sanguinetti,
A.~Sciab\`a,
G.~Sguazzoni,
R.~Tenchini,$^{1}$
A.~Venturi,
P.G.~Verdini
\samepage
\begin{center}
\parbox{15.5cm}{\sl\samepage
Dipartimento di Fisica dell'Universit\`a, INFN Sezione di Pisa,
e Scuola Normale Superiore, I-56010 Pisa, Italy}
\end{center}\end{sloppypar}
\vspace{2mm}
\begin{sloppypar}
\noindent
G.A.~Blair,
G.~Cowan,
M.G.~Green,
T.~Medcalf,
J.A.~Strong,
\mbox{J.H.~von~Wimmersperg-Toeller}
\nopagebreak
\begin{center}
\parbox{15.5cm}{\sl\samepage
Department of Physics, Royal Holloway \& Bedford New College,
University of London, Surrey TW20 OEX, United Kingdom$^{10}$}
\end{center}\end{sloppypar}
\vspace{2mm}
\begin{sloppypar}
\noindent
R.W.~Clifft,
T.R.~Edgecock,
P.R.~Norton,
I.R.~Tomalin
\nopagebreak
\begin{center}
\parbox{15.5cm}{\sl\samepage
Particle Physics Dept., Rutherford Appleton Laboratory,
Chilton, Didcot, Oxon OX11 OQX, United Kingdom$^{10}$}
\end{center}\end{sloppypar}
\vspace{2mm}
\begin{sloppypar}
\noindent
\mbox{B.~Bloch-Devaux},$^{1}$
P.~Colas,
S.~Emery,
W.~Kozanecki,
E.~Lan\c{c}on,
\mbox{M.-C.~Lemaire},
E.~Locci,
P.~Perez,
J.~Rander,
\mbox{J.-F.~Renardy},
A.~Roussarie,
\mbox{J.-P.~Schuller},
J.~Schwindling,
A.~Trabelsi,$^{21}$
B.~Vallage
\nopagebreak
\begin{center}
\parbox{15.5cm}{\sl\samepage
CEA, DAPNIA/Service de Physique des Particules,
CE-Saclay, F-91191 Gif-sur-Yvette Cedex, France$^{17}$}
\end{center}\end{sloppypar}
\vspace{2mm}
\begin{sloppypar}
\noindent
S.N.~Black,
J.H.~Dann,
R.P.~Johnson,
H.Y.~Kim,
N.~Konstantinidis,
A.M.~Litke,
M.A. McNeil,
\linebreak
G.~Taylor
\nopagebreak
\begin{center}
\parbox{15.5cm}{\sl\samepage
Institute for Particle Physics, University of California at
Santa Cruz, Santa Cruz, CA 95064, USA$^{13}$}
\end{center}\end{sloppypar}
\vspace{2mm}
\begin{sloppypar}
\noindent
C.N.~Booth,
S.~Cartwright,
F.~Combley,
M.~Lehto,
L.F.~Thompson
\nopagebreak
\begin{center}
\parbox{15.5cm}{\sl\samepage
Department of Physics, University of Sheffield, Sheffield S3 7RH,
United Kingdom$^{10}$}
\end{center}\end{sloppypar}
\vspace{2mm}
\begin{sloppypar}
\noindent
K.~Affholderbach,
A.~B\"ohrer,
S.~Brandt,
C.~Grupen,$^{1}$
A.~Misiejuk,
G.~Prange,
U.~Sieler
\nopagebreak
\begin{center}
\parbox{15.5cm}{\sl\samepage
Fachbereich Physik, Universit\"at Siegen, D-57068 Siegen,
 Germany$^{16}$}
\end{center}\end{sloppypar}
\vspace{2mm}
\begin{sloppypar}
\noindent
G.~Giannini,
B.~Gobbo
\nopagebreak
\begin{center}
\parbox{15.5cm}{\sl\samepage
Dipartimento di Fisica, Universit\`a di Trieste e INFN Sezione di Trieste,
I-34127 Trieste, Italy}
\end{center}\end{sloppypar}
\vspace{2mm}
\begin{sloppypar}
\noindent
J.~Rothberg,
S.~Wasserbaech
\nopagebreak
\begin{center}
\parbox{15.5cm}{\sl\samepage
Experimental Elementary Particle Physics, University of Washington, Seattle, 
WA 98195 U.S.A.}
\end{center}\end{sloppypar}
\vspace{2mm}
\begin{sloppypar}
\noindent
S.R.~Armstrong,
K.~Cranmer,
P.~Elmer,
D.P.S.~Ferguson,
Y.~Gao,
S.~Gonz\'{a}lez,
O.J.~Hayes,
H.~Hu,
S.~Jin,
J.~Kile,
P.A.~McNamara III,
J.~Nielsen,
W.~Orejudos,
Y.B.~Pan,
Y.~Saadi,
I.J.~Scott,
J.~Walsh,
Sau~Lan~Wu,
X.~Wu,
G.~Zobernig
\nopagebreak
\begin{center}
\parbox{15.5cm}{\sl\samepage
Department of Physics, University of Wisconsin, Madison, WI 53706,
USA$^{11}$}
\end{center}\end{sloppypar}
}
\footnotetext[1]{Also at CERN, 1211 Geneva 23, Switzerland.}
\footnotetext[2]{Now at Universit\'e de Lausanne, 1015 Lausanne, Switzerland.}
\footnotetext[3]{Also at Dipartimento di Fisica di Catania and INFN Sezione di
 Catania, 95129 Catania, Italy.}
\footnotetext[4]{Also Istituto di Fisica Generale, Universit\`{a} di
Torino, 10125 Torino, Italy.}
\footnotetext[5]{Also Istituto di Cosmo-Geofisica del C.N.R., Torino,
Italy.}
\footnotetext[6]{Supported by the Commission of the European Communities,
contract ERBFMBICT982894.}
\footnotetext[7]{Supported by CICYT, Spain.}
\footnotetext[8]{Supported by the National Science Foundation of China.}
\footnotetext[9]{Supported by the Danish Natural Science Research Council.}
\footnotetext[10]{Supported by the UK Particle Physics and Astronomy Research
Council.}
\footnotetext[11]{Supported by the US Department of Energy, grant
DE-FG0295-ER40896.}
\footnotetext[12]{Now at Departement de Physique Corpusculaire, Universit\'e de
Gen\`eve, 1211 Gen\`eve 4, Switzerland.}
\footnotetext[13]{Supported by the US Department of Energy,
grant DE-FG03-92ER40689.}
\footnotetext[14]{Also at Rutherford Appleton Laboratory, Chilton, Didcot, UK.}
\footnotetext[15]{Permanent address: Universitat de Barcelona, 08208 Barcelona,
Spain.}
\footnotetext[16]{Supported by the Bundesministerium f\"ur Bildung,
Wissenschaft, Forschung und Technologie, Germany.}
\footnotetext[17]{Supported by the Direction des Sciences de la
Mati\`ere, C.E.A.}
\footnotetext[18]{Supported by the Austrian Ministry for Science and Transport.}
\footnotetext[19]{Now at SAP AG, 69185 Walldorf, Germany.}
\footnotetext[20]{Now at Harvard University, Cambridge, MA 02138, U.S.A.}
\footnotetext[21]{Now at D\'epartement de Physique, Facult\'e des Sciences de Tunis, 1060 Le Belv\'ed\`ere, Tunisia.}
\footnotetext[22]{Now at Department of Physics, Ohio State University, Columbus, OH 43210-1106, U.S.A.}
\footnotetext[23]{Also at Dipartimento di Fisica e Tecnologie Relative, Universit\`a di Palermo, Palermo, Italy.}
\footnotetext[24]{Now at CERN, 1211 Geneva 23, Switzerland.}
%
\setlength{\parskip}{\saveparskip}
\setlength{\textheight}{\savetextheight}
\setlength{\topmargin}{\savetopmargin}
\setlength{\textwidth}{\savetextwidth}
\setlength{\oddsidemargin}{\saveoddsidemargin}
\setlength{\topsep}{\savetopsep}
\normalsize
\newpage
\pagestyle{plain}
\setcounter{page}{1}

\newpage
\setcounter{page}{1}
\pagestyle{plain}

\section{Introduction}

In the Standard Model CP violation arises from a complex phase of the quark
mixing matrix~\cite{KM}, and this can accommodate the observed CP violation
in the K sector~\cite{PDG}.  Precise predictions can be made of relations
between asymmetries expected in B decays, and their detailed study will
provide an important test of the model.  The first step, however, is to
establish the existence of CP violation in B decays.  This has been
attempted using inclusive methods, where the expected asymmetry is small,
${\cal O}(10^{-3})$, beyond the sensitivity of current
experiments~\cite{inclusive}.  The alternative is to use exclusive decays,
where the asymmetries are predicted to be large, ${\cal O}(1)$, but the
branching ratios are small.

The decay $\rm B^0 \rightarrow J/\psi\,K^0_S$ is known as the gold-plated
mode for such studies, due to its clean experimental signature and low
theoretical uncertainty. The final state is a CP eigenstate, to which both
$\rm B^0$ and $\rm \overline{B}{}^0$ can decay.  The interference between
their direct and indirect decays via $\rm B^0$--$\rm \overline{B}{}^0$ mixing
leads to a time-dependent CP asymmetry given by
\begin{equation}
A(t) \equiv \frac{\rm \Gamma (B^0 \rightarrow J/\psi\,K^0_S) -
\Gamma (\overline{B}{}^0 \rightarrow J/\psi\,K^0_S)}
{\rm \Gamma (B^0 \rightarrow J/\psi\,K^0_S) +
\Gamma (\overline{B}{}^0 \rightarrow J/\psi\,K^0_S)} 
= -\sin 2\beta\,\sin \Delta m_{\rm d}\/t \ .
\end{equation}
Here $\rm \Gamma (B^0 \rightarrow J/\psi\,K^0_S)$ represents the rate of
particles that were {\em produced} as $\rm B^0$ decaying to $\rm
J/\psi\,K^0_S$ at proper time $t$, $\Delta m_{\rm d}$ is the oscillation
frequency of the $\rm B^0$, and $\beta$ is an angle of the ``unitarity
triangle'' of the quark mixing matrix, given by the following combination
of matrix elements: $\arg(-V_{\rm cd}V^*_{\rm cb}/V_{\rm td}V^*_{\rm
tb})$~\cite{SM}. Information can be obtained indirectly about the value of
$\sin 2\beta$, within the context of the Standard Model, from the
combination of other measurements that constrain the matrix elements, such
as $\Delta m_{\rm d}$, charmless B decays and CP violation in the K sector.
Many such fits have been made, typically preferring large positive values
for $\sin 2\beta$ in the range 0.4--0.8~\cite{triangle}; a recent example
gave $\sin 2\beta = 0.75 \pm 0.09$~\cite{Mele}.

The first published attempt at a direct measurement was made by
OPAL~\cite{OPAL}. They selected 24 candidates with an estimated purity of
60\% and reported a value outside the physical region, $\sin 2\beta = 3.2\
^{+1.8}_{-2.0} \pm 0.5$, which was nevertheless interpreted as favouring
large positive values.  CDF published an analysis based on 395 candidates
with a purity of about 40\%, although half of the sample has poor
proper-time determination~\cite{CDF}. They measured $\sin 2\beta = 0.79\
^{+0.41}_{-0.44}$ (statistical and systematic errors combined).

The key to making such a measurement at LEP is to keep the efficiency as
high as possible.  Using the latest branching ratio ${\cal B} {\rm
(B^0\rightarrow J/\psi\,K^0) = (8.9 \pm 1.2)\times 10^{-4}}$~\cite{PDG}, in
the complete dataset of ALEPH, about 30 signal events are expected before
the reconstruction efficiency is applied.  The production state of the $\rm
B^0$ must also be determined (or ``tagged'').  The precision on $\sin
2\beta$ scales as $1/(1-2w)$, where $w$ is the mistag rate, the fraction of
incorrectly tagged events.  Using a neural-network technique to combine the
information from many observables, the lowest possible mistag rate is aimed
for, whilst providing a tag for every event.

In this paper, after a brief description of the ALEPH detector, the event
selection is discussed.  Details are given of the proper-time measurement,
and the production-state tagging.  The unbinned likelihood fit for the
asymmetry is then presented, followed by a discussion of checks and
systematic uncertainties.

\section{Detector}

A detailed description of the ALEPH detector can be found
in~\cite{detector} and its performance in~\cite{performance}.  Charged
particles are tracked in a two-layer silicon vertex detector with
double-sided readout ($r$--$\phi$ and $z$), surrounded by a cylindrical
drift chamber and a large time projection chamber (TPC), together measuring
up to 33 space points along the trajectory. These detectors are immersed in
a 1.5\,T axial magnetic field, providing a transverse momentum resolution
of $\Delta p/p = (6\times 10^{-4})\,p$ at high momentum (for $p$ in
GeV$/c$) and a three-dimensional impact parameter resolution of $25\,\mu$m.
The TPC also allows particle identification to be performed through the
measurement of specific ionization ($dE/dx$). A finely segmented
electromagnetic calorimeter of lead/wire-chamber sandwich construction
surrounds the TPC.  Estimators $R_{\rm T}$, $R_{\rm L}$ and $R_{\rm I}$ are
formed for electron identification, for the transverse and longitudinal
shower shape in the calorimeter and for the $dE/dx$ in the TPC, respectively;
they are calculated as the difference between the measured and expected
value for electrons, divided by the expected uncertainty. The iron return
yoke of the magnet is instrumented with streamer tubes to form a hadron
calorimeter and is surrounded by two additional double layers of streamer
tubes to aid muon identification.

The LEP1 data were recently reprocessed using improved reconstruction
algorithms. A new pattern recognition algorithm for the vertex detector
allows groups of nearby tracks to be analysed together, searching for hit
assignments that minimize the overall $\chi^2$ of the event.  Information
from the wires and pads of the TPC are also combined to improve the spatial
and $dE/dx$ resolution~\cite{lifetime}.

Monte Carlo simulated events are used to study both the signal and the
background.  The simulation is based on JETSET~\cite{JETSET} and is
described in detail in~\cite{MC}.  To tune the selection cuts for
background suppression, 6.5~million hadronic Monte Carlo events are used,
corresponding to about 1.5 times the data statistics.  In addition, a large
sample of signal Monte Carlo events is used for the determination of the
expected signal mass distribution, reconstruction efficiency, and for
training of the neural network for tagging.

\section{Event selection}

Data taken by ALEPH in the years 1991--95 at the Z resonance are used,
corresponding to 4.2 million hadronic Z decays. 
Hadronic events are selected in the data as described in~\cite{hadronic}.
The production vertex position is reconstructed on an event-by-event basis
using the constraint of the average beam-spot position~\cite{performance}.

First, the $\rm J/\psi\rightarrow\ell^+\ell^-$ reconstruction is performed.
The daughter tracks are required to have momentum greater than 2\,GeV$/c$,
distance of closest approach to the primary vertex of less than 2\,cm
transverse to and 10\,cm along the beam axis, four or more hits in the TPC,
and polar angle satisfying $|\cos\theta\, |<0.95$. All oppositely-charged
pairs of such tracks, with opening angle satisfying $\cos\theta_{\ell\ell}
> 0.85$, are investigated for lepton identification. They must both be
identified as muons or electrons using loose identification criteria.
For muons, a pattern of hits in the HCAL consistent with a muon is
required~\cite{leptons}; for electrons, cuts are made on the estimators:
$|R_{\rm T}|<4$, $R_{\rm L}>-3$ and $R_{\rm I}>-4$. The invariant mass of
the lepton pair is required to be in the range 2.6--3.3 GeV$/c^2$, with the
large window around the $\rm J/\psi$ mass (particularly on the low side)
maintaining high efficiency in the presence of radiative decays or
bremsstrahlung.

\begin{figure}[t]
\vspace*{-5mm} 
\begin{center}\vspace*{-2.5cm}\epsfig{file=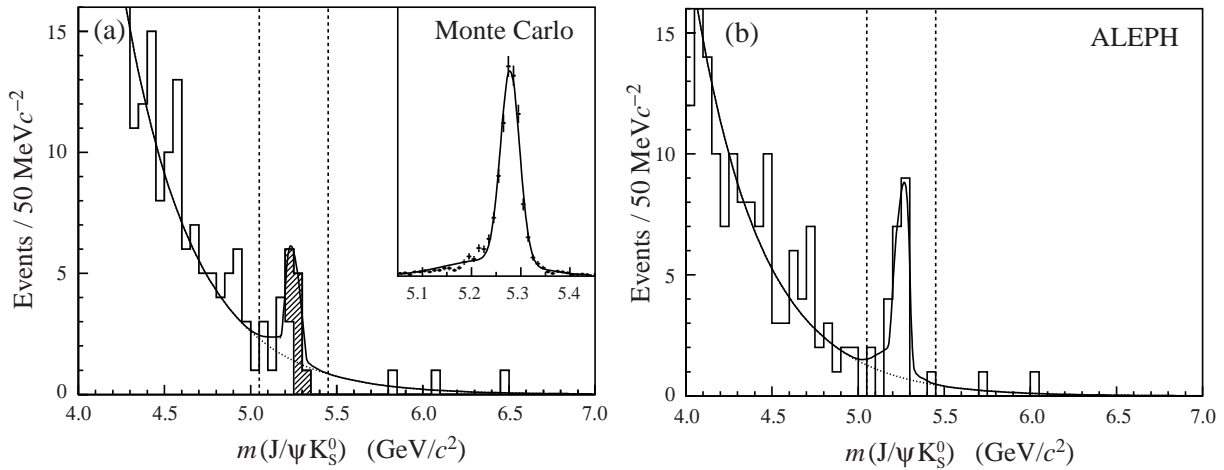,height=9.5cm}\end{center}
\vspace*{-5mm} 
\caption{\small Reconstructed $\rm J/\psi\, K^0_S$ mass (a) for hadronic Monte
  Carlo, (b) for the data; the superimposed fits are described in the text,
  and the dashed lines indicate the signal region. In (a) the events that
  truly originated from a $\rm B^0 \rightarrow J/\psi\, K^0_S$ decay are
  shaded, and the insert shows the mass distribution for a large sample of
  signal Monte Carlo events.
\label{mass}}\end{figure}

Next the $\rm K^0_S\rightarrow\pi^+\pi^-$ reconstruction is performed, as
described in~\cite{V0}. The distance of closest approach of the two
daughters must be less than 5\,mm, and the $dE/dx$ for each daughter is
required to be within three standard deviations of the expected value for a
pion. The angle between the reconstructed directions of the $\rm K^0_S$ and
$\rm J/\psi$ is required to satisfy $\cos\theta_{\rm\psi K} > 0.85$, and
the resultant of their momenta must satisfy $p_{\rm B}>22$\,GeV$/c$.  The
reconstructed $\rm K^0_S$ invariant mass is required to be within
15\,MeV$/c^2$ of its nominal value.

A fit is made for the $\rm B^0$ decay vertex, using the two lepton tracks
and the $\rm K^0_S$. The vertex is required to be successfully
reconstructed, and the decay length is determined from the distance between
the production and decay vertices, projected along the momentum vector of
the $\rm B^0$ candidate.  The decay length is required to be
greater than $-1$\,mm, and its calculated error less than
1\,mm. Finally, the $\rm J/\psi\, K^0_S$ invariant mass is calculated,
assigning the nominal masses to the two particles. The resulting
distribution is shown as an insert in Fig.~\ref{mass}\,(a) for signal Monte
Carlo. The peak is fitted with the sum of two Gaussian functions, the first
accounting for 72\% of the events with a width of 20\,MeV$/c^2$ and the
second with width 90\,MeV$/c^2$, offset to lower mass by 50\,MeV$/c^2$.  The
signal region is defined as 5.05--5.45\,GeV$/c^2$, and the overall
reconstruction efficiency is 28\%.

When the event selection is applied to the hadronic Monte Carlo sample, the
resulting $\rm J/\psi\, K^0_S$ mass distribution is shown in
Fig~\ref{mass}\,(a).  There are 20 events in the signal region, of which 11
are from background. The superimposed fit is the sum of an exponential
shape to describe the background and the signal shape discussed above.  The
four fitted parameters are the background slope and normalization, and the
signal mass and normalization.

The mass distribution that results when the event selection is applied to
the data is shown in Fig.~\ref{mass}\,(b).  A clear signal is seen for the
$\rm B^0$, with 23 events in the signal region.  A fit similar to that of
the hadronic Monte Carlo is made, giving a fitted $\rm B^0$ mass of $5.26
\pm 0.01$\,GeV$/c^2$, slightly lower than, but consistent with, the
world-average value of $m_{\rm B}=5.279$\,GeV$/c^2$. The fit is used to
assign an event-by-event background probability for each event in the
signal region, which is used in the maximum likelihood fit for the CP
asymmetry.  Their sum corresponds to 6.6 background events, giving an
average background fraction $f_{\rm bkg}=0.29\pm 0.06$, where the
uncertainty is estimated by varying the parametrizations within their
statistical errors, and using alternative shapes for the background.  The
fitted shape and normalization of the background are consistent with those
seen in the Monte Carlo.  

\begin{figure}[t]
\vspace*{0cm} 
\begin{center}\vspace*{1cm}\epsfig{file=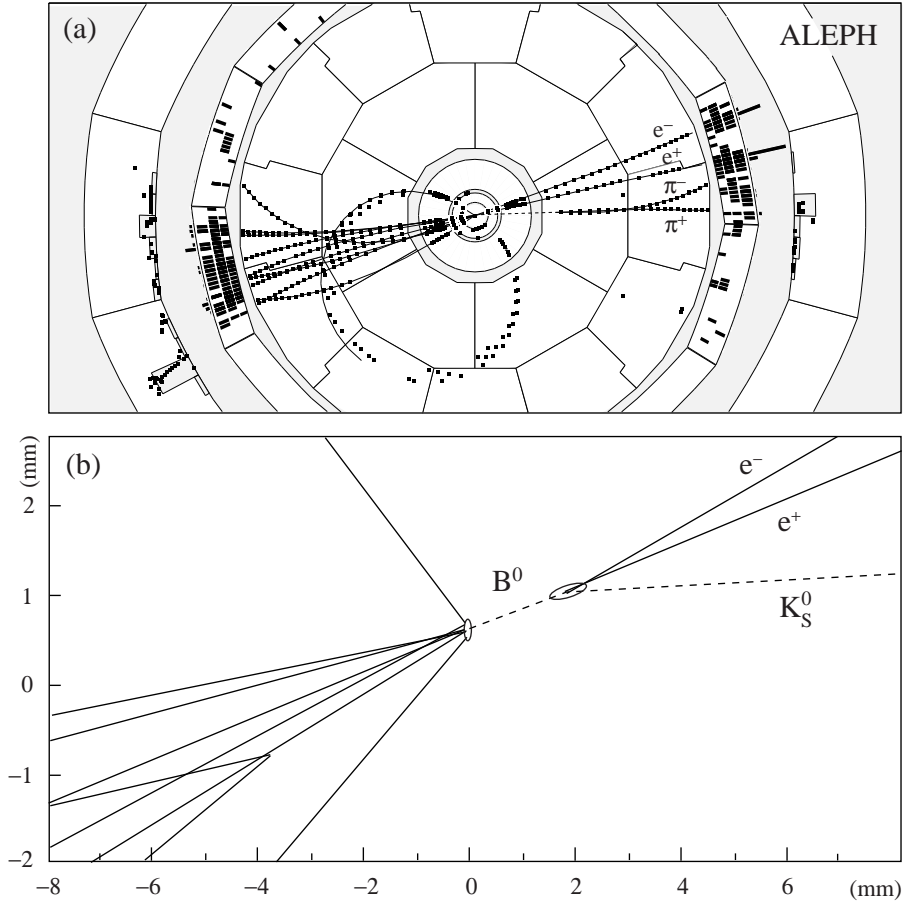,height=12cm}\end{center}
\vspace*{-3mm}
\caption{\small Event display of a $\rm B^0\rightarrow J/\psi\,K^0_S$ candidate in the
  data, with $\rm J/\psi \rightarrow e^+e^-$: (a) fish-eye view of the
  projection transverse to the beam axis, (b) zoom into the vertex region for
  the same projection, with reconstructed vertices marked with ellipses,
  and the reconstructed neutrals marked with dashed lines.
\label{event}}\end{figure}

After subtraction of the background, about 16 signal events remain, to be
compared with the predicted number of $9\pm 2$, calculated from the
expected production rate and decay branching ratios. Figure~\ref{event} shows
a particularly clean signal candidate, in which there are no other charged
tracks in the signal hemisphere.

\section{Proper-time determination}

The decay length of the $\rm B^0$ is determined from vertex reconstruction
as described in the previous section.  The decay-length resolution
determined using the Monte Carlo simulation is reasonably described by a
Gaussian distribution of width $200\,\mu$m, although there are small tails
from events with poorly measured vertices.  The uncertainty on the decay
length is estimated by propagating the production and decay vertex errors,
which are calculated in turn from the errors of the tracks used in their
determination. The pull distribution, given by the difference of true and
reconstructed decay lengths divided by the calculated decay-length
uncertainty, is close to being normally distributed. It has a fitted
Gaussian width of 1.1, indicating that the event-by-event estimate of
decay-length uncertainty is reasonably accurate.

As the signal events are fully reconstructed, the momentum resolution is
excellent.  There is a Gaussian core of 1\% relative error, with an overall
RMS of 2.5\%.  This small uncertainty on the momentum measurement gives a
significant contribution to the overall proper-time uncertainty only for
events with long decay lengths, greater than about 1\,cm.

The proper time is given by $t=d_{\rm B} m_{\rm B} / p_{\rm B}$. Its
uncertainty is calculated as follows:
\begin{equation}
\sigma_t = t\ \sqrt{\left(\frac{\sigma_d}{d_{\rm B}}\right)^2 + 
  \left(\frac{\sigma_p}{\raisebox{.3ex}{$p_{\rm B}$}}\right)^2}\ .
\end{equation}
The first term in parentheses is taken from the event-by-event measurement
of the decay length $d_{\rm B}$ and its error $\sigma_d$, scaled by a
factor $1.1\pm 0.2$, where the uncertainty is taken into account for
systematic studies; the second term is taken as $(2\pm 2)$\%. Despite being
conservative, these estimates of the uncertainty lead to a negligible
effect on the measured asymmetry, as the characteristic scale of the
proper-time development of the asymmetry is $1/\Delta m_{\rm d}\sim 2$\,ps,
much longer than the typical proper-time resolution of 0.1\,ps.

Of the 11 background events in the signal region for the hadronic Monte
Carlo, nine involve tracks from b-hadron decays. The apparent proper time and
its error are determined for these background events, following the
procedure discussed above, and a fit made for their effective lifetime
$\tau_{\rm bkg}$. The probability density function is taken as the
convolution of an exponential lifetime distribution with a Gaussian
resolution function, and the fit gives \mbox{$\tau_{\rm bkg}=1.4\
^{+0.5}_{-0.3}$\,ps}.

\section{Production-state tagging}

The extraction of the CP asymmetry from the signal events requires the
determination of whether they originated from a ${\rm B^0}$ or $\rm
\overline{B}{}^0$ at production.  This is achieved by studying the
properties of the rest of the event, excluding the lepton and pion pairs
that come from the signal $\rm B^0$ decay.  For this purpose, two
hemispheres are defined with respect to the thrust axis of the event,
determined using both charged and neutral energy-flow
objects~\cite{performance}.  These are used to separate information from
the same and opposite sides of the event with respect to the ${\rm B^0}$
candidate.  Properties of the opposite side are used to determine the
particle/antiparticle nature of the other b hadron that was produced in
conjunction with the signal ${\rm B^0}$, and thus indirectly determine its
production state.  The same side carries information from the fragmentation
process that produced the signal ${\rm B^0}$, which can also be used in the
production-state determination.

The calculation of the tag for the opposite side starts with the search for
a secondary vertex due to b-hadron decay.  This is achieved using a
topological vertexing algorithm that combines information from all charged
tracks in the hemisphere.  Jets are reconstructed from the charged tracks
and neutral energy-flow objects using the JADE algorithm~\cite{JADE}, with
a jet-resolution parameter of 0.02.  The highest energy jet and the jet
which forms the highest invariant mass with it are selected, and the
secondary vertex is constrained to lie along the direction of the selected
jet in its hemisphere, within errors. Each track is then assigned a
relative probability ${\cal P}_{\rm v}$ that it originates from the
secondary vertex.

Next, the b-hadron flight direction is estimated.  Jets are reconstructed
with a jet-resolution parameter lowered to 0.0044, as this gives an
improved estimate of the b-hadron direction~\cite{leptons}. If more than
one jet is found in the hemisphere considered, the b-jet candidate is
chosen on the basis of the kinematic properties of its tracks and the
presence of lepton candidates. The leading track of the b-jet candidate is
then used as a seed for a second cone-based jet algorithm~\cite{BTCONE},
which is taken as a first estimate of the b-hadron flight direction.  In
two-jet events, the thrust axis is chosen instead. The uncertainties on the
reconstructed angles are parametrized from the simulation, as a function of
the jet momentum.  A second estimate of the direction is taken as the
vector joining the primary and secondary vertices, and its error is
parametrized as a function of the measured decay length.  The two estimates
are averaged using their parametrized errors, and the result taken as the
b-hadron flight direction.

\begin{figure}[t]\vspace*{0cm}
\begin{center}\epsfig{file=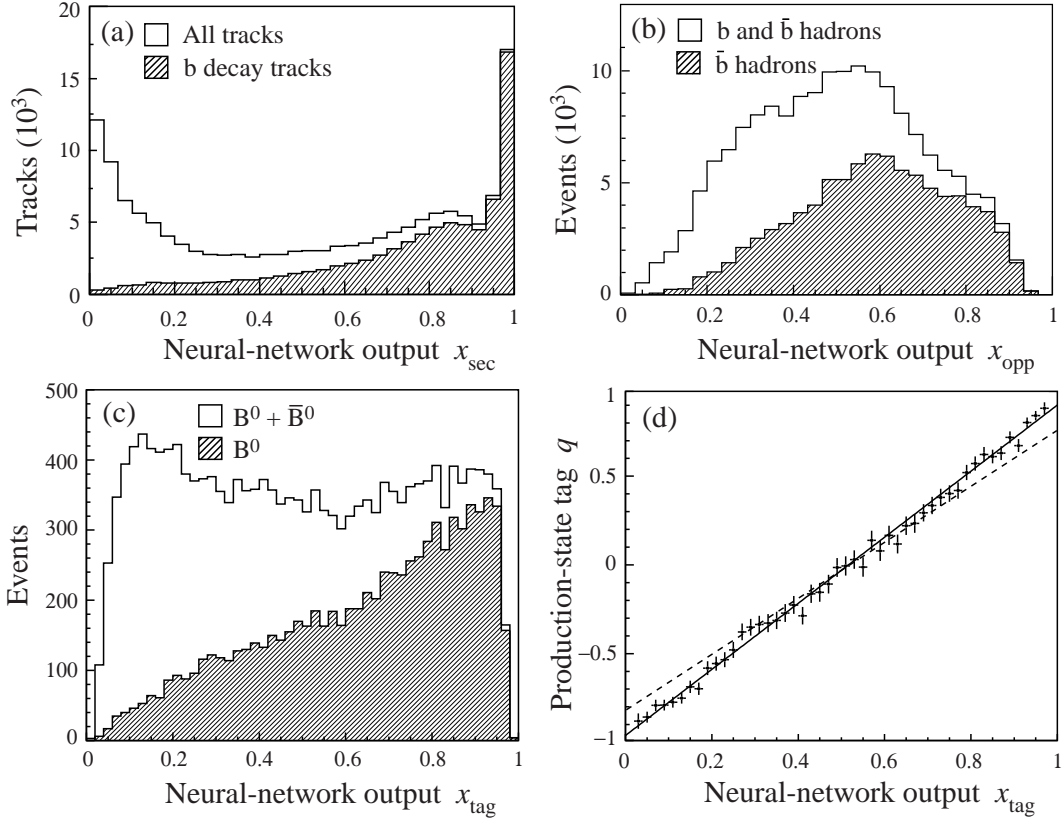,height=11cm}\end{center}\vspace*{-5mm}
\caption{\small Neural-network output distributions for signal Monte Carlo
  events: (a)~separation of b-decay and fragmentation tracks, $x_{\rm
  sec}$, with the contribution from b-decay tracks shaded;
  (b)~opposite-side production state, $x_{\rm opp}$, with the
  contribution from hemispheres containing $\rm \overline{b}$ hadrons
  shaded; (c)~overall event production state, $x_{\rm tag}$, with the
  contribution from $\rm B^0$ decays shaded.  (d)~Calibration of the
  production-state tag $q$ versus the neural-network output, with
  superimposed linear fit; the effect of degrading the mistag rate for
  systematic studies is indicated by the dashed line.
\label{tag}}\end{figure}

To construct the charge estimators, b-decay and fragmentation tracks must
be distinguished.  This is achieved by combining three variables for each
track in the hemisphere: the rapidity and longitudinal momentum relative to
the b-hadron flight direction, and the vertex assignment probability ${\cal
P}_{\rm v}$.  They are combined in a neural network, along with a
hemisphere-based indicator of the quality of the inclusive vertex finding,
which acts as a ``control variable''. This does not directly discriminate
between the b-decay and fragmentation tracks, but improves the overall
performance of the neural network as the discriminating power of the
other inputs vary as a function of the control variable.  The single track
separation output is shown in Fig.~\ref{tag}\,(a), using neural networks
trained separately for classes of tracks passing different quality
criteria.  The neural-network output value $x_{\rm sec}$ is converted to a
probability ${\cal P}_{\rm sec}$ that the track comes from the secondary
vertex.  The b-hadron momentum is then determined from the sum of charged
energy assigned to the secondary vertex (based on ${\cal P}_{\rm sec}$
weights), the longitudinal fraction of neutral energy in the jet, and the
missing energy in the hemisphere.

Nine charge-sensitive inputs are used for the opposite-side
production-state tag:
\begin{enumerate}

\item{\em Jet charge ($\kappa=0.5$):} 
\begin{equation}
  Q_{\rm J}=\frac{\sum_i q_i (\mbox{\boldmath $p$}_i\cdot 
  \mbox{\boldmath $d$}_{\rm J})^\kappa} {\sum_i (\mbox{\boldmath $p$}_i\cdot 
  \mbox{\boldmath $d$}_{\rm J})^\kappa}\ ,
\end{equation}
where $i$ runs over the charged tracks in the hemisphere, $q_i$ and
{\boldmath $p$}$_i$ are the charge and momentum of the track,
{\boldmath $d$}$_{\rm J}$ is the b-hadron flight direction.

\item{\em Jet charge ($\kappa=1.0$):} defined as in Eq.~3.

\item{\em Charge of the track with highest longitudinal momentum} in the
hemisphere, calculated relative to the b-hadron flight direction.

\item{\em Total charge:} $Q_{\rm tot} =
\sum_{i} \, q_i$ where $i$ runs over the charged tracks
in the b jet.

\item{\em Secondary vertex charge:} ${Q_{\rm vtx}} =
\sum_i \, {\cal P}_{\rm sec}^i \, q_i$ where $i$ runs over all
charged tracks in the hemisphere.

\item{\em Weighted primary vertex charge:}
\begin{equation}
{Q_{\rm pri}} = \frac{\sum_i ( 1 -  {\cal P}_{\rm sec}^i
) \, q_i \, (\mbox{\boldmath $p$}_i\cdot 
  \mbox{\boldmath $d$}_{\rm J})^\kappa}{ \sum_i ( 1 -  {\cal P}_{\rm sec}^i
) \, (\mbox{\boldmath $p$}_i\cdot 
  \mbox{\boldmath $d$}_{\rm J})^\kappa}\ ,
\end{equation}
where $i$ runs over all charged tracks in the hemisphere and $\kappa=1.0$.

\item{\em Weighted secondary vertex charge:} as for $Q_{\rm pri}$ but
  replacing $(1-{\cal P}_{\rm sec}^i)$ with ${\cal P}_{\rm sec}^i$, and
  taking $\kappa=0.3$.

\item{\em Decay kaon charge:} kaons are identified using $dE/dx$, based
upon the ratio of probabilities that the measured ionization is due to a
kaon, relative to either a kaon or a pion.  This variable is combined with
the track momentum, longitudinal momentum and ${\cal P}_{\rm sec}$ value
to select kaons from the $\rm b \rightarrow c \rightarrow s$ cascade using
a further neural network.  The reconstructed b-hadron momentum in the
hemisphere is also used as a control variable.  The charge of the track
with the highest output from the neural network is used to sign the output
value.

\item{\em Lepton charge:} lepton identification is performed with tighter
requirements than those described previously~\cite{lifetime}.  If more than
one lepton candidate is selected, that with the highest $p_{\rm T}$ with
respect to the jet axis is used.  The lepton transverse momentum is signed
by the track charge.

\end{enumerate}

Finally, six additional control variables are used: $\cos \, \theta_{\rm
thrust}$, the charged track multiplicity, the spread in the measured values
of ${\cal P}_{\rm sec}$: $\sigma_{\rm sec} = \sum_i {\cal P}_{\rm sec}
\left( 1 - {\cal P}_{\rm sec} \right)$, the reconstructed b-hadron
momentum, the reconstructed decay length, and the lepton momentum (if a
lepton has been selected).  These are combined along with the charge
estimators using a neural network, which takes into account correlations
between the variables. It provides a single output, $x_{\rm opp}$, which is
shown in Fig.~\ref{tag}\,(b).

Information on the production state from the same hemisphere as the $\rm
B^0$ candidate is limited to the charged tracks from fragmentation.
Excluding the tracks from the $\rm B^0$, seven charge estimators are
constructed using tracks coming from the primary vertex in the signal
hemisphere.  In a similar way to the opposite-side analysis, these include
two jet charges with $\kappa$ values of 0.5 and 1.0, the charge of the
track with the highest longitudinal momentum with respect to the jet, one
momentum-weighted and two longitudinal-momentum-weighted primary vertex
charges with $\kappa$ values of 1.0, 0.3 and 1.0, respectively, and finally
the sum of track charges inside the jet.  The output of the opposite-side
neural network, $x_{\rm opp}$, is then combined with these same-side charge
estimators and the following control variables: $\sigma_{\rm sec}$, $\cos
\theta_{\rm thrust}$, and the same-side charged track multiplicity, using
another neural network.  The output of this event tag, $x_{\rm tag}$, is
shown in Fig.~\ref{tag}\,(c).

If events with $x_{\rm tag} > 0.5$ are taken to have a ${\rm B^0}$ in the
production state and those with $x_{\rm tag} < 0.5$ to have a $\rm
\overline{B}{}^0$, the fraction of incorrect tags (the average mistag rate)
is 27.8\%.  The event-by-event value of the neural-network output $x_{\rm
tag}$ is used in the measurement of the asymmetry. Due to its peaked
distribution, seen in Fig.~\ref{tag}\,(c), the effective mistag rate is
lower than the average value quoted above.  Using an overlap integral
technique~\cite{discrim} the effective mistag rate is found to be 24.1\%,
measured with an independent sample of simulated events, with an expected
difference of $(0.2\pm 0.2)$\% between the effective mistag values for $\rm
B^0$ and $\rm \overline{B}{}^0$ hemispheres. The opposite-side tag alone
gives average and effective mistag rates of 31.4\% and 27.6\%, respectively.

Finally, the production-state tag $q$ is calculated from the neural-network
output $x_{\rm tag}$, correcting for the purity in each bin of
Fig.~\ref{tag}\,(c) using
\begin{equation}
q = \frac{2\,F(x_{\rm tag})}{F(x_{\rm tag})+G(x_{\rm tag})} - 1\ ,
\end{equation}
where $F(x_{\rm tag})$ is the distribution of $x_{\rm tag}$ for events
produced as $\rm B^0$, shaded in the figure, and $G(x_{\rm tag})$ is the
distribution for $\rm \overline{B}{}^0$; $q=+1$ for events produced as $\rm
B^0$ and $-1$ for $\rm \overline{B}{}^0$ in the case of perfect tagging.
The relationship between $x_{\rm tag}$ and $q$, determined with signal
Monte Carlo events, is consistent with linearity, as shown in
Fig.~\ref{tag}\,(d). It is parametrized as a straight line 
$q = a_{\rm tag} + 2\,b_{\rm tag}(x_{\rm tag}-0.5)$, with fitted coefficients 
$a_{\rm tag}=-0.01\pm 0.01$ and $b_{\rm tag}=0.95\pm 0.01$.

\section{Asymmetry measurement}

\begin{figure}[t]
\begin{center}\vspace*{-1cm}
\epsfig{file=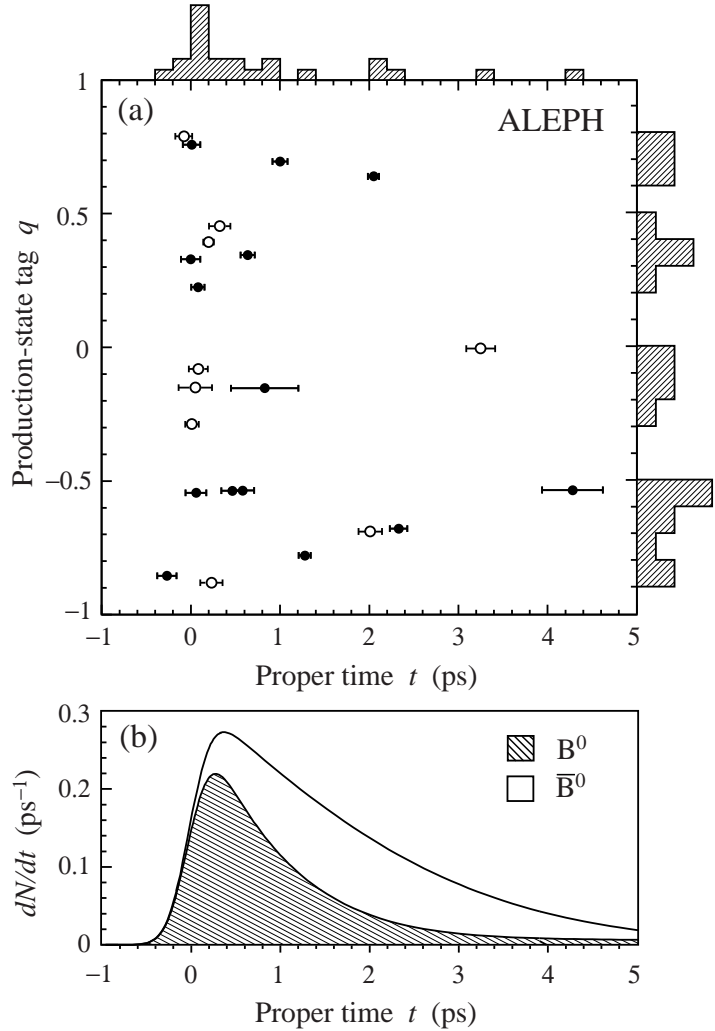,height=15cm}\end{center}\vspace*{-1.8cm}
\caption{\small (a) Production-state tag $q$ versus proper time $t$
  for the 23 events in the signal region of the data. The proper-time
  uncertainty is indicated for each point, and histograms of the two
  variables are shaded.  The solid points indicate events with a background
  probability less than 30\%. (b)~Expected proper-time distribution of $\rm
  B^0$ and $\rm \overline{B}{}^0$ events, for fixed values of the CP
  asymmetry and proper-time resolution: $\sin 2\beta=0.7$,
  $\sigma_t=0.2$\,ps.
\label{qt}}\end{figure}

The production-state tag $q$ is plotted against the proper time $t$ in
Fig.~\ref{qt}\,(a), for the 23 events in the signal region of the data.
Considering the events which have a clear tag result, $|q|>0.5$, there are
4 ``$\rm B^0$-like'' events (positive $q$) and 9 ``$\rm
\overline{B}{}^0$-like''.  Furthermore, the excess of events with negative
$q$ is more noticeable with increasing proper time, as would be expected for a
negative CP asymmetry.

To extract a measurement of the asymmetry, an unbinned likelihood is
calculated. The probability density function expected for the signal
distribution is given by
\begin{equation}
  {\cal P}_{\rm sig}(t,q) = \frac{e^{-t/\tau_{\rm d}}}{2\,\tau_{\rm
  d}}(1-q\,\sin 2\beta\, \sin\Delta m_{\rm d}\/t)\ .
\end{equation}
The $\rm B^0$ lifetime $\tau_{\rm d} = 1.548 \pm 0.032$\,ps and 
$\Delta m_{\rm d}=0.472 \pm 0.017$\,ps$^{-1}$ are fixed to the
central values of their world averages~\cite{PDG} (the uncertainties are
taken into account in systematic error studies).  

A convolution is made of this signal distribution and a Gaussian resolution
function ${\cal R}(\sigma_t)$ with width given by the event-by-event
proper-time resolution calculated in Section~4. 
The result is illustrated in Fig.~\ref{qt}\,(b), where the
contributions from $\rm B^0$ and $\rm \overline{B}{}^0$ are indicated as a
function of proper time for fixed values of the asymmetry and resolution.

The probability density function for background events ${\cal P}_{\rm
bkg}(t, q)$ is taken to have the same form as that of the signal but
replacing $\tau_{\rm d}\rightarrow\tau_{\rm bkg}$ and $\sin
2\beta\rightarrow a_{\rm bkg}$, where the effective lifetime of the
background $\tau_{\rm bkg}$ was determined in Section~4. The background
asymmetry $a_{\rm bkg}$ is taken to be zero, but is varied to study
possible systematic effects. The likelihood of an event is then calculated
as
\begin{equation}
{\cal L}_i = 
(1-f_{\rm bkg})\,{\cal P}_{\rm sig}(t,q)\otimes {\cal R}(\sigma_t) +
f_{\rm bkg}\,{\cal P}_{\rm bkg}(t,q)\otimes {\cal R}(\sigma_t)\ ,
\end{equation}
where $f_{\rm bkg}$ is the event-by-event background fraction discussed in
Section~3.  A generalized likelihood function is used, adding a
Poisson term to the combined likelihoods of the signal candidates:
\begin{equation}
L = \frac{\nu^n\,e^{-\nu}}{n!} \prod_{i=1}^{n} {\cal L}_i\ ,
\end{equation}
where $n=23$ is the number of observed candidates, and $\nu$ is the number
expected.

\begin{figure}[t]
\begin{center}\vspace*{-25mm}\epsfig{file=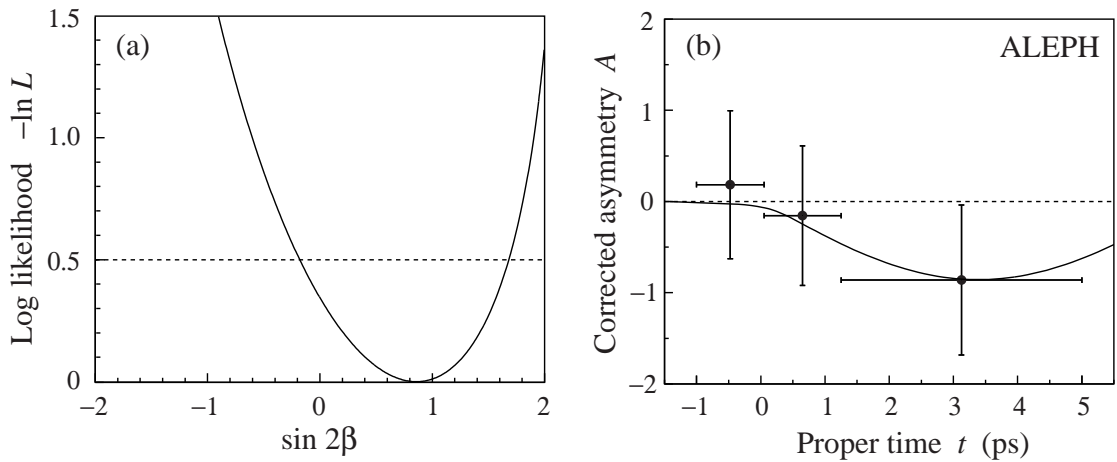,height=8cm}\end{center}
\vspace*{-6mm} 
\caption{\small (a)~Log-likelihood versus $\sin 2\beta$ from the fit, with
  respect to the minimum. (b)~Reconstructed asymmetry versus proper time,
  after correction for 
  the average dilution in each time bin. The result of the fit
  for the asymmetry is shown superimposed; note that this
  is not fitted directly to the binned data shown.
\label{asym}}\end{figure}

A fit is made with two free parameters, $\sin 2\beta$ and $\nu$, with the
results $\sin 2\beta = 0.84\ ^{+0.82}_{-1.04}$, $\nu=23.3\ ^{+5.1}_{-4.5}$.
A scan of the log-likelihood versus $\sin 2\beta$ is shown in
Fig.~\ref{asym}\,(a).  The result of the measurement is displayed as a
function of proper time in Fig.~\ref{asym}\,(b). For the purpose of display
the data are divided into three proper-time bins, and the average asymmetry
is corrected for the average dilution (from mistagging and background) in
each bin. Negative asymmetry, and thus positive $\sin 2\beta$, is favoured.

\section{Checks and systematic uncertainties}

The statistical error from the fit has been checked using a fast Monte Carlo
simulation.  Event samples the same size as the data were generated
according to Eq.~6 and 7, using the resolution function and tagging
distributions discussed above. The value of $\sin 2\beta$ was set to that
measured for the data, and all other parameters were fixed to their nominal
values in the fit. The number of background events within the sample
was varied according to Poisson statistics.  The CP asymmetry of each
sample was measured as if it were data, and the central value and errors
recorded. This was repeated for many samples. The values reconstructed for
the positive and negative errors of the data are consistent with the
distributions of values seen in the simulated samples.  Furthermore, the
measured errors reasonably estimate the spread of the measured central
value about its true input value: the pull is normally distributed.

\begin{table}[t]
\centering
\vspace*{-5mm}
\caption{\small Contributions to the systematic uncertainty on $\sin 2\beta$.}
\vspace*{3mm}
\begin{tabular}{|l|l|r|}
\hline
Source & Variation & $\sigma_{\rm sys}$ \\
\hline
$\rm B^0$ lifetime & $\tau_{\rm d} = 1.548 \pm 0.032$\,ps & $<0.01$ \\
$\rm B^0$ oscillation & $\Delta m_{\rm d} = 0.472 \pm 0.017$\,ps$^{-1}$ &
0.04 \\
Decay-length resolution & $\sigma_d \times (1.1\pm 0.2)$ & $<0.01$ \\
Momentum resolution & $\Delta p/p=0.02\pm 0.02$ & $<0.01$ \\
Background level & $f_{\rm bkg}=0.29\pm 0.06$ & 0.09 \\
Background lifetime & $\tau_{\rm bkg}=1.4\ ^{+0.5}_{-0.3}$\,ps & 0.01 \\
Background asymmetry & $a_{\rm bkg}=0.0\pm 0.2$ & 0.04 \\
Tag calibration offset & $a_{\rm tag}=-0.01\pm 0.02$ & 0.10 \\
Tag calibration slope & $b_{\rm tag}=0.95\pm 0.05$ & 0.05 \\
\hline
Total & & 0.16 \\
\hline
\end{tabular}
\end{table}

The contributions to the systematic error are listed in Table~1. The
lifetime and oscillation frequency of the signal were varied within their
world-average uncertainties, and the effect on the measured asymmetry taken
as a systematic error.  For the lifetime, a check was made by fitting for the
lifetime of selected signal Monte Carlo events, giving $\tau =1.56\pm
0.02$\,ps, in agreement with the input value of $1.56$\,ps, indicating no
evidence for a systematic bias.  Similarly their asymmetry was measured to be
$-0.02\pm 0.03$, in agreement with the input value of zero. The
decay-length and momentum resolution parametrizations were varied as
discussed in Section~4.

For the background, the uncertainty on the level was discussed in
Section~3.  The effective lifetime of the background was varied within the
large uncertainty measured using the hadronic Monte Carlo, given in
Section~4.  To increase the statistics for the study of the CP asymmetry of
the background, events in the side-band region $4.0<m({\rm J/\psi\,
K_S^0})<5.0$\,GeV$/c^2$ were investigated. There are 140 such events in the
data, with asymmetry $0.09\pm 0.23$, indicating no significant effect.
About one event is expected in the signal region from the decay $\rm
B^0\rightarrow J/\psi\,K^{*0}$ with $\rm K^{*0}\rightarrow K^0_S\,\pi^0$,
predicted to be mainly CP even.  This would correspond to a background
asymmetry of less than 0.1.  For the systematic error evaluation the
asymmetry of the background, $a_{\rm bkg}$, was varied by $\pm 0.2$.

Uncertainties arising from the production-state tag can be considered as
being due to knowledge of the overall mistag rate for signal events in data
and the possible difference in the individual mistag rates for $\rm B^0$
and $\rm \overline{B}{}^0$ events.  Samples of events were isolated in data
with a similar topology to that of the signal, where there is a b hadron of
known charge with decay products that can be cleanly separated from
fragmentation tracks.  Three samples of $\rm B^+$ candidates were selected
for this purpose. The same neural network was used for tagging their
production state as for $\rm B^0\rightarrow J/\psi\,K^0_S$, except that the
sign of the same-side charge estimators is reversed.  This is required as a
$\overline{\rm b}$ quark combines with a d (u) quark to produce a $\rm B^0$
($\rm B^+$), and the accompanying $\overline{\rm d}$ ($\overline{\rm u}$)
quarks have opposite sign.

\begin{figure}[t]\vspace*{0cm}
\begin{center}\epsfig{file=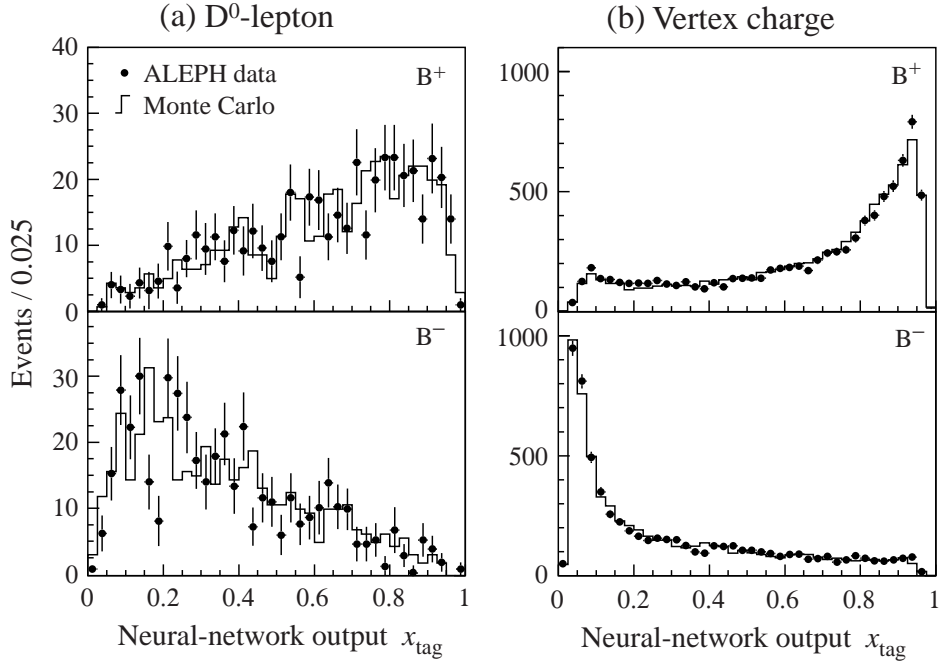,height=9cm}\end{center}\vspace*{-3mm}
\caption{\small Production-state tag neural-network output for (a)~$\rm
  \overline{D}{}^0\ell^+$ and (b)~vertex-charge selected samples.  In each
  figure the upper plot is for events selected as $\rm B^+$ candidates, the
  lower for $\rm B^-$.
\label{sys}}\end{figure}

\begin{enumerate}
\item{\em $\rm B^+ \rightarrow J/\psi\, K^+$ decays:} The signal selection
is modified slightly, to select $\rm B^+\rightarrow J/\psi\,K^+$ decays
(and charge conjugate).  The reconstructed $\rm J/\psi\,K^+$ mass in the
data is shown in Fig.~\ref{charged}\,(a); a clear signal is seen, with 52
events in the signal region compared to an expected background of about
17. In this channel the efficiency is similar to that of $\rm
J/\psi\,K^0_S$, but the product branching ratio is about three times
higher.  Now the charge of the kaon indicates whether the signal was
produced by $\rm B^+$ or $\rm B^-$, and the resulting distributions of the
neural-network output $x_{\rm tag}$ are compared in
Fig.~\ref{charged}\,(b). From a study of side-band events, the background
is found to have a similar tagging distribution to the signal.  The average
mistag rate is 26\% in signal Monte Carlo and is measured to be $(27\pm
6)$\% in the data, giving confidence that the Monte Carlo simulation
reproduces faithfully the features of the data used for tagging.  The
asymmetry and lifetime have also been measured for these events, with the
results $0.09\pm 0.41$ and $1.57\ ^{+0.23}_{-0.20}$\,ps, respectively.  The
latter agrees with the world-average value of $1.653\pm
0.028$\,ps~\cite{PDG}.

\begin{figure}[t]
\vspace*{-1cm}  
\begin{center}\vspace*{10mm}\epsfig{file=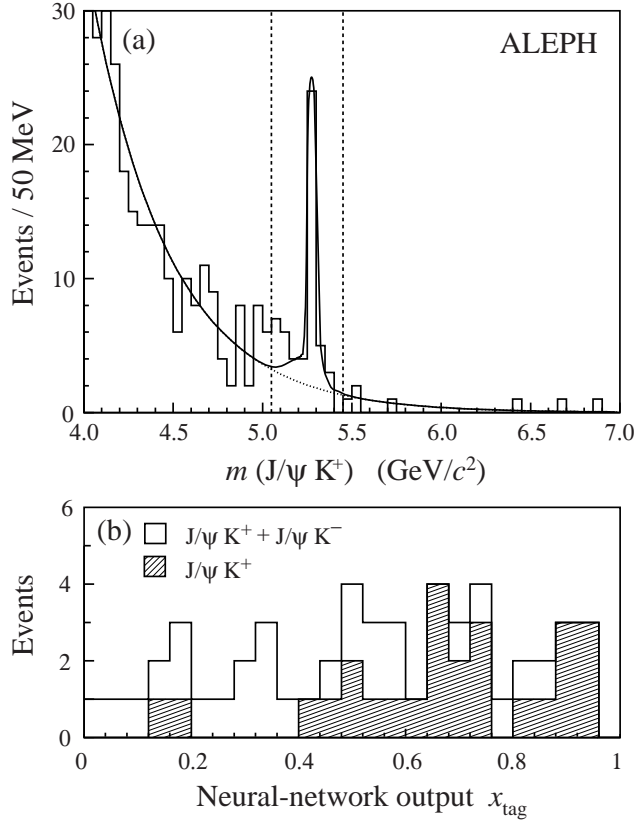,height=11cm}\end{center}
\vspace*{-5mm}  
\caption{\small (a)~Reconstructed $\rm J/\psi\, K^+$ mass, for the data.
  (b) Production-state tag neural-network output $x_{\rm tag}$, when
  applied to events from the signal region.
\label{charged}}\end{figure}

\item{\em ${\rm B}^+ \rightarrow \overline{\rm D}{}^0 \ell^+ X$ decays:} A
selection is made of $\rm D^0$ candidates, decaying to a kaon and a pion of
opposite charge.  Only those $\rm D^0$ candidates are used which are found
to lie in the same hemisphere as an identified lepton (e or $\mu$) with
transverse momentum greater than 0.5\,GeV$/c$ and charge opposite in sign
to that of the pion from the $\rm D^0$ decay.  The estimated b purity of
the sample is 87\%, dominated by $\rm B^\pm$ decays.  The lepton charge
indicates whether the signal was produced by a $\rm B^+$ or $\rm B^-$, and
the resulting tag distributions are shown in Fig.~\ref{sys}\,(a).  The
effective mistag rate determined from data is $(24.5\pm 0.6)$\%, with a
Monte Carlo expectation of $(24.1\pm 0.5)$\%.

\item{\em Inclusive $\rm B^\pm$ decays using vertex charge:} A selection of
$\rm B^\pm$ decays is made by requiring a hemisphere with $\sigma_{\rm sec}
< 0.32$ opposite to a b-tagged hemisphere~\cite{btag}, that gives an
estimated b purity of 95\%.  A cut is then made on the absolute value of
the vertex charge $|Q_{\rm vtx}| > 0.6$ in the selected hemisphere, which
isolates a sample containing an expected 73\% of $\rm B^\pm$ decays.  The
sign of the vertex charge indicates whether the signal is produced by a
$\rm B^+$ or a $\rm B^-$ and the resulting distributions of the tag output
are shown in Fig.~\ref{sys}(b).  Although significant differences are seen
in this sample between the tag distributions for $\rm B^+$ and $\rm B^-$
candidates, due to material effects, they are well reproduced by the Monte
Carlo. The effective mistag rate determined from data is $(21.9\pm 0.2)$\%,
with a Monte Carlo expectation of $(21.0\pm 0.1)$\%.

\end{enumerate}

The largest data-Monte Carlo disagreement for the mistag rates is 
$(0.9\pm 0.2)$\% from the vertex-charge selected sample.  In addition,
separating the samples into $\rm B^+$ and $\rm B^-$ decays as shown in
Fig.~\ref{sys} and determining the mistag rates separately for each results
in a maximum observed discrepancy of $(2.4\pm 1.3)$\% from the $\rm
\overline{D}{}^0 \ell^+$ selected events.  These are taken as systematic
uncertainties for the production-state mistag rate of the $\rm B^0
\rightarrow J/\psi\, K^0_S$ signal. Their effect is propagated to the fit
for the CP asymmetry by modifying the calibration of the tag, the
parametrization of $q$ versus $x_{\rm tag}$. A degraded mistag rate leads
to a reduced slope of the calibration, as illustrated in
Fig.~\ref{tag}\,(d).  Similarly, a $\rm B^0$--$\rm \overline{B}{}^0$
tagging difference is seen as an offset to the calibration. The variations
applied to the parametrization are listed in Table~1.

Adding the systematic error contributions in quadrature, the final
measurement is
\begin{equation}
\sin 2\beta = 0.84\ ^{+0.82}_{-1.04}\pm 0.16\ ,
\end{equation}
where the first error is statistical and the second systematic. 

\section{Conclusion}

\begin{figure}[t]\vspace*{0cm}
\begin{center}\epsfig{file=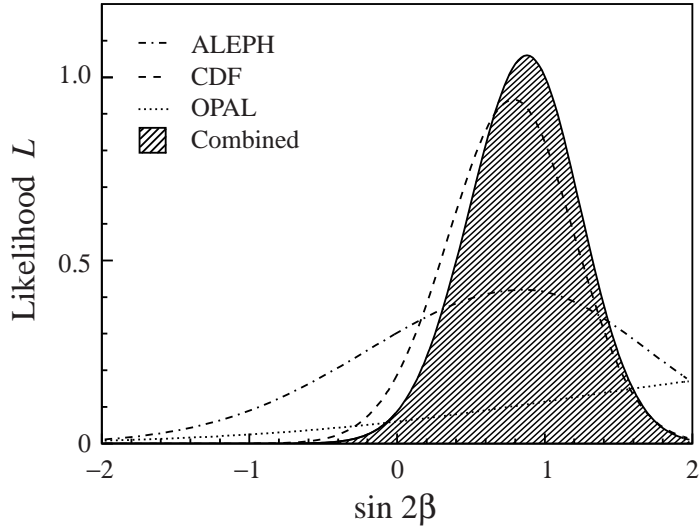,height=7cm}\end{center}\vspace*{-5mm}
\caption{\small Likelihood distributions versus $\sin 2\beta$ for the three
  existing measurements (dashed lines) and their combination (shaded).
\label{lik}}\end{figure}

An analysis of $\rm B^0\rightarrow J/\psi\,K^0_S$ decays has been
performed, with $\rm J/\psi \rightarrow e^+e^-$ or $\mu^+\mu^-$ and $\rm
K^0_S \rightarrow \pi^+\pi^-$. A reconstruction
efficiency of 28\% is achieved. From the full dataset of ALEPH at LEP1 of
4.2~million hadronic Z decays, 23~candidates are selected with an estimated
purity of 71\%.  They are used to measure the CP asymmetry of this decay,
with the result $\sin 2\beta = 0.84\ ^{+0.84}_{-1.05}$,
where the uncertainty is dominated by the statistics. 

This result is compared with the other two published measurements in
Fig.~\ref{lik}. The likelihood of each measurement is approximated as
a Gaussian distribution on either side of the central value, with width equal
to the sum in quadrature of statistical and systematic errors.  The three
log-likelihoods can be summed to give a combined result, i.e.\ neglecting any
correlation between the systematic errors of the different experiments
(expected to be small). The resulting likelihood is shown by the shaded
distribution in the figure and corresponds to $\sin 2\beta = 0.88\
^{+0.36}_{-0.39}$.  The integral of the likelihood for $\sin 2\beta>0$ is
77\% for the ALEPH result alone, or 67\% if the total integral is limited to
the physical region $|{ \sin 2\beta }|<1$.  The corresponding fractions are
98.6\% and 98.0\% for the combination of the three analyses. Thus 
the confidence level that CP violation has been seen in this channel is
increased to 98\%, compared to 93\% for CDF alone~\cite{CDF}.

Preliminary results from the BABAR and BELLE experiments~\cite{osaka} are
consistent with the combined result given above, with similar precision and
slightly lower central values.

\section*{Acknowledgements}

It is a pleasure to thank the CERN accelerator divisions for the successful
operation of LEP.  We are also grateful to the engineers and technicians in our
institutes for their contribution to the performance of ALEPH.  Those of us
from non-member states thank CERN for its hospitality.

\end{document}